\documentclass[reprint, superscriptaddress, aps]{revtex4-2}

\usepackage{graphicx}
\usepackage{dcolumn}
\usepackage[version=4]{mhchem}
\usepackage{hyperref}

\newcommand{\Oeighteen}{\ce{^{18}O}\;}
\newcommand{\Tett}{\ce{^{132}Te}\;}
\newcommand{\Teth}{\ce{^{130}Te}\;}

\begin{document}

\title{Lifetime of the \ce{{4}^{+}_{1}} state of \ce{^{132}Te}}
	
\author{H.~Mayr}
\email{hmayr@ikp.tu-darmstadt.de}
\author{T.~Stetz}
\affiliation{Technische Universität Darmstadt, Department of Physics, Institute for Nuclear Physics, Schlossgartenstraße 9, 64289 Darmstadt, Germany}
\author{V.~Werner}
\affiliation{Technische Universität Darmstadt, Department of Physics, Institute for Nuclear Physics, Schlossgartenstraße 9, 64289 Darmstadt, Germany}
\affiliation{Helmholtz Forschungsakademie Hessen für FAIR (HFHF), Campus Darmstadt, Schlossgartenstraße 2, 64289 Darmstadt, Germany}
\author{N.~Pietralla}
\affiliation{Technische Universität Darmstadt, Department of Physics, Institute for Nuclear Physics, Schlossgartenstraße 9, 64289 Darmstadt, Germany}
\author{M.~Beckers}
\author{A.~Blazhev}
\author{A.~Esmaylzadeh}
\author{J.~Fischer}
\author{R.-B.~Gerst}
\affiliation{Universität zu Köln, Institute for Nuclear Physics, Zülpicher Straße 77, 50937 Köln, Germany}
\author{K.A.~Gladnishki}
\affiliation{University of Sofia St. Kliment Ohridski, Faculty of Physics, 5 James Bourchier blvd., 1164 Sofia, Bulgaria}
\author{K.E.~Ide}
\affiliation{Technische Universität Darmstadt, Department of Physics, Institute for Nuclear Physics, Schlossgartenstraße 9, 64289 Darmstadt, Germany}
\author{J.~Jolie}
\affiliation{Universität zu Köln, Institute for Nuclear Physics, Zülpicher Straße 77, 50937 Köln, Germany}	
\author{V.~Karayonchev}
\affiliation{Tri University Meson Facility TRIUMF, 4004 Wesbrook Mall, Vancouver, BC V6T 2A3, Canada}
\author{E.~Kleis}
\author{H.~Kleis}
\author{P.~Koch}
\affiliation{Universität zu Köln, Institute for Nuclear Physics, Zülpicher Straße 77, 50937 Köln, Germany}	
\author{D.~Kocheva}
\affiliation{University of Sofia St. Kliment Ohridski, Faculty of Physics, 5 James Bourchier blvd., 1164 Sofia, Bulgaria}
\author{C.M.~Nickel}
\affiliation{Technische Universität Darmstadt, Department of Physics, Institute for Nuclear Physics, Schlossgartenstraße 9, 64289 Darmstadt, Germany}
\author{T.~Otsuka}
\affiliation{Center for Nuclear Study, University of Tokyo, Hongo, Bunkyo-ku, Tokyo 113-0033, Japan}
\author{A.~Pfeil}
\affiliation{Universität zu Köln, Institute for Nuclear Physics, Zülpicher Straße 77, 50937 Köln, Germany}
\author{G.~Rainovski}
\affiliation{University of Sofia St. Kliment Ohridski, Faculty of Physics, 5 James Bourchier blvd., 1164 Sofia, Bulgaria}
\author{F.~von~Spee}
\affiliation{Universität zu Köln, Institute for Nuclear Physics, Zülpicher Straße 77, 50937 Köln, Germany}
\author{M.~Stoyanova}
\affiliation{University of Sofia St. Kliment Ohridski, Faculty of Physics, 5 James Bourchier blvd., 1164 Sofia, Bulgaria}
\author{Y.~Tsunoda}
\affiliation{Center for Nuclear Study, University of Tokyo, Hongo, Bunkyo-ku, Tokyo 113-0033, Japan}
\affiliation{Center for Computational Sciences, University of Tsukuba, 1-1-1 Tennodai, Tsukuba 305-8577, Japan}
\author{R.~Zidarova}
\affiliation{Technische Universität Darmstadt, Department of Physics, Institute for Nuclear Physics, Schlossgartenstraße 9, 64289 Darmstadt, Germany}

\begin{abstract}
    The evolution of the collectivity of tellurium isotopes from mid-shell towards $N=82$ is currently based mainly on properties of the first excited $2^+$ states. To extend structural information in this isotopic chain, in particular with respect to the balance of microscopic, seniority-type and collective excitations, electric quadrupole transition strengths from $4^+$ states need to be considered. An experiment was performed to determine the $4_1^+$ lifetime of \Tett via the recoil-distance Doppler-shift method at the University of Cologne tandem accelerator. The isotope of interest was populated in the two neutron-transfer reaction $^{130}$Te($^{18}$O,$^{16}$O)$^{132}$Te$^*$. The $E2$ decay transition strength has been determined to be $B(E2; 4^+_1\rightarrow 2^+_1) = 9.3(10)\, \text{W.u.}$ This value differs from literature data but compares favourably to shell model calculations.
\end{abstract}
\maketitle

\section{Introduction}
Isotopes in the vicinity of double shell closures offer insight into the single-particle structure of the corresponding near-spherical isotopes. These are also the regions where the nuclear shell model is best applicable, and model valence spaces, residual interactions, effective charges and $g$ factors can be tailored to the simplest proton and neutron configurations identified in such nuclei. Even-even isotopes offer insight into the basic couplings of valence protons and neutrons in the active valence orbitals, forming the lowest-lying excited states of these nuclei. In particular, nuclei along a proton- or neutron-shell closure allow to identify the most basic proton and neutron configurations. 

Seniority-type excitations \cite{TALMI19711} occur where a pair of protons or neutrons in a respective $l_j$ orbital couples to the resulting spin of an eigenstate in the form $\rho(l_j^2)^{(J)}$, $\rho\in[\pi,\nu]$. Isotopes with only one pair of valence protons and neutrons, each, are then the simplest systems where such proton and neutron pairs can, again pairwise, couple to low angular momenta, for example as $[\pi(l_j^2)^{(J_\pi)} \nu(l_j^2)^{(J_\nu)}]^{(J)}$ or more complicated configurations where the proton and neutron pairs may break up and scatter across multiple orbitals. The proton-neutron interaction then merges both, proton and neutron pairs, into low-lying nuclear eigenstates. Therefore, basic seniority-type excitations can be seen as building blocks in the formation of proton-neutron coupled states \cite{heydesau,PhysRevC.71.054304,wex,holt,casp}, which evolve into more collective states for nuclides located further into the open shell. In isotopes near double shell closures, both types of configurations, seniority-type or multiple coupled pairs, can be in competition and coexist in the low-lying level scheme.

Key observables for the investigation of the onset and evolution of quadrupole collectivity are $E2$ transition strengths. For seniority-dominated states, only the $\Delta s=2$ seniority-changing $2^+_1 \rightarrow 0^+_1$ $E2$ transition is allowed, whereas the $\Delta s=0$ $E2$ transitions between the higher-lying states of the yrast band are suppressed \cite{TALMI19711,Ressler04}. For multiple coupled-pair configurations, the latter transitions gain in $E2$ strengths. Hence, the interplay between both will directly imprint on the $E2$ transition rate within the ground-state band, in particular, the $B(E2;4^+_1 \rightarrow 2^+_1)$ value of interest in this work.

\begin{figure}[t]
    \includegraphics[width=\linewidth]{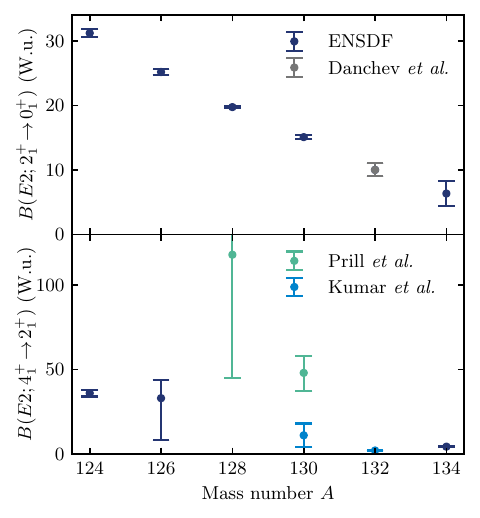}
    \caption{The evolution of the $B(E2; 2^+_1\rightarrow 0^+_1)$ values (upper panel)  in the tellurium isotopes follows a clear trend, declining from $N=72$ towards the shell closure at $N=82$. A similar trend cannot be seen for the $B(E2; 4^+_1\rightarrow 2^+_1)$ values due to the insufficient number of precision data (lower panel). Weighted averages of the $B(E2; 2_1^+\rightarrow0_1^+)$ values were taken from the Nuclear Data Sheets $A=126$--$130,134$ \cite{A126, A128, A130, A134}. For \Tett the $B(E2; 2_1^+\rightarrow0_1^+)$ value from Danchev \textit{et al.}~\cite{Danchev} was used. The $B(E2; 4_1^+\rightarrow2_1^+)$ values are taken from Kumar \textit{et al.}~\cite{Kumar}, Prill \textit{et al.}~\cite{prill} and Refs.~\cite{Saxena024316,STOKSTAD1967507, A134}.} The uncertainty of the $^{128}$Te value is cut off for visibility.
    \label{fig:b(e2)_evol}
\end{figure}

The $B(E2;2_1^+\rightarrow0_1^+)$ values from the (typically) first-excited states are well known for the tellurium isotopes around $N=78$ \cite{A126, A128, A130, A134, Danchev}. They follow a nearly linear decline from $N=72$ towards the shell closure at $N=82$ as can be seen in the top panel of Fig. \ref{fig:b(e2)_evol}. The $4^+_1$ member of the ground-state band, and its $B(E2;4_1^+\rightarrow2_1^+)$ value give more insight into the degree of collectivity of an even-even nucleus. The current literature values for the tellurium isotopic chain towards $N=82$ are included at the bottom of Fig. \ref{fig:b(e2)_evol} \cite{Saxena024316,STOKSTAD1967507, prill, Kumar, A134}. A clear trend cannot be seen for the $B(E2;4^+_1\rightarrow 2^+_1)$ values due to the small number of data points with sufficient accuracy and precision. It is the purpose of this paper to report the measurement of the $4^+_1$ lifetime of \Tett. It also provides the ratio
\begin{equation}
\label{eq:b42}
B_{4/2} = \frac{B(E2;4_1^+\rightarrow2_1^+)}{B(E2;2_1^+\rightarrow0_1^+)}\,,
\end{equation}
which serves as an indicator for the degree of collectivity. The $B_{4/2}$ ratio takes benchmark values in the collective limits of a spherical harmonic vibrator with $B^{\rm vib}_{4/2} = 2$, or quadrupole-deformed limits with $B^{\rm rot}_{4/2} = 1.4$, regardless of axial symmetry. For near-magic nuclei with seniority-dominated low-lying states, $B_{4/2}$ can take values even lower than unity. Hence, for $^{132}$Te with its proximity to the $Z=50$ and $N=82$ shell closures this quantity will serve as an indicator of the balance between spherical-collective and spherical-seniority dominated structures. Furthermore, the $B_{4/2}$ ratio is more sensitive to the nuclear wave functions than its energy correspondent, the $R_{4/2} = E(4^+_1)/E(2^+_1)$ ratio with benchmark values of $R^{\rm vib}_{4/2} = 2$ for collective vibrators or $R^{\rm rot}_{4/2} = 3.33$ for well-deformed rotors, and $R_{4/2}$ significantly smaller than 2 only in the seniority-dominated regime. For the tellurium isotopes approaching the $N=82$ shell closure, $R_{4/2}(^{130}$Te$) = 1.95$ \cite{A130} indicates a rather spherical-collective structure, whereas $R_{4/2}(^{132}$Te$) = 1.72$ \cite{A132} indicates a significant role of non-collective contributions to the wave functions of the respective states. This work is concerned with a more sensitive test of the wave functions through the measurement of the corresponding $E2$ strength.

A claim for the lifetime of the $4_1^+$ state of \Tett has recently been published with $\tau(4_1^+)=61(5)\,$ps \cite{Kumar} from a $\gamma\gamma$ fast-timing analysis. In that experiment by Kumar \textit{et al.\!} excited states of \Tett were populated via $\beta^-$ decay of \ce{^{132}Sb} produced through thermal neutron-induced fission. Another study by Roberts \textit{et al.\!} \cite{Roberts}, also using the fast-timing technique but the \Teth($^{7}$Li,$\alpha p$) incomplete-fusion transfer reaction as a population mechanism, resulted in an upper limit for the lifetime of $\tau(4_1^+)\leq 58\,$ps. Assuming that the value by Kumar {\it et al.}, located at the edge of the upper limit by Roberts {\it et al.} is correct, one obtains a $B_{4/2}$ ratio of only $B_{4/2} = 0.20(3)$, clearly below the collective limit and in line with a dominantly seniority-conserving transition. This value would be situated even below the corresponding value of the neutron-magic $^{134}$Te.
Recent results on the neighbouring isotope $^{130}$Te, however, showed a stark disagreement between a value by Kumar {\it et al.} \cite{Kumar} and one by Prill {\it et al.} \cite{prill} (see Fig. \ref{fig:b(e2)_evol}), raising doubts on the claimed accuracy of the literature on the $4^+_1$ lifetime of $^{132}$Te. This doubt is further supported by the fact that the reported lifetime by Kumar {\it et al.} is close to the sensitivity limit of the fast-timing technique which was utilised there.

In order to resolve this situation, within the present work a direct lifetime measurement of the $4_1^+$ state of \Tett was conducted.
In view of the $4_1^+$ lifetimes of neighbouring tellurium isotopes, the expected lifetime of the $4_1^+$ state of \Tett can be estimated to be in the picosecond range. Hence, the recoil-distance Doppler-shift (RDDS) method \cite{DEWALD2012786} is applicable and was employed to measure the lifetime of interest, ruling out its present literature value, as will be discussed in the following.

\section{Experimental Details}
\begin{figure*}[t]
    \centering
    \includegraphics[width=\linewidth]{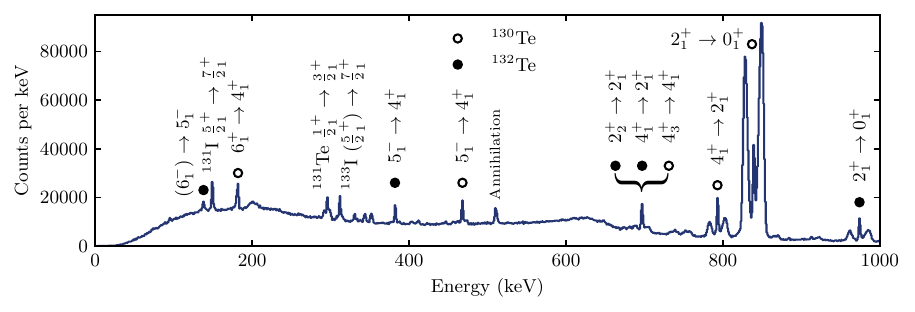}
    \caption{The observed $\gamma$-ray energy spectrum is shown with applied particle-coincidence condition as displayed in the inset of Fig. \ref{fig:spectrum_697}. The spectrum is summed over all detectors and all distance settings. Therefore, some transitions show Doppler-shifted behaviour and appear as triple-peak structures. The peaks of the most prominent transitions are marked. The $4_1^+\to2_1^+$ transition of \Tett contributes the most to the peak marked by the bracket. The contamination of the $2_2^+\to2_1^+$ transition of \Tett is visible in the enlarged spectrum depicted in Fig. \ref{fig:spectrum_697}. The possible contamination by the $4_3^+\to4_1^+$ transition of \Teth is discussed in section \ref{sec:data_analysis}. Spectroscopic information is taken from \cite{A130, A131, A132, A133}.}
    \label{fig:spectrum_overview}
\end{figure*}

The experiment was conducted at the 10-MV FN-tandem accelerator at the University of Cologne. An \Oeighteen beam impinged on a $0.9\,$mg/cm$^2$ thick \Teth (enrichment $\geq$97\,\%) target with a beam energy of approximately 61.5 MeV, populating states of interest of \Tett by the $^{130}$Te($^{18}$O, $^{16}$O)$^{132}$Te$^*$ two-neutron transfer reaction. The \Teth material was backed by $0.4\,$mg/cm$^2$ natural vanadium with the \Teth layer downstream. The stopper foil consisted of $2.5\,$mg/cm$^2$ thick natural magnesium, sufficient to stop the recoiling tellurium ions.

The $\gamma$ radiation was detected by 11 high-purity germanium (HPGe) detectors. They were mounted in two rings and passively shielded with lead. Five of the HPGe detectors were set up at a backward angle of 142$^\circ$ and six at a forward angle of 45$^\circ$ with respect to the beam axis. Six solar cells were mounted at backward angles approximately $11\,$cm away from the target for particle detection of beam-like fragments covering an angle of approximately $120^\circ$-$165^\circ$. The combination of $\gamma$-ray and particle detectors allows to set coincidence conditions to select transitions of target-like nuclei. Fig.~\ref{fig:spectrum_overview} shows the obtained $\gamma$-ray energy spectrum of the whole detector array summed over all distance settings. Applying a particle-energy coincidence condition on the back-scattered ejectiles is necessary since fusion-evaporation cross sections are about one order of magnitude larger than the two-neutron transfer cross sections at beam energies around the Coulomb barrier. Events from fusion-evaporation reactions can be suppressed by exploiting the difference in reaction kinematics between the transfer and fusion-evaporation reactions as can be seen in the inset of Fig.~\ref{fig:spectrum_697}. Additionally, the solar cells at backward angles restrict the kinematics of the recoiling reaction products.

In an RDDS experiment the nucleus of interest is produced and its excited states populated within a thin target foil. Deexcitation of the nucleus can either occur in flight or after being stopped in a stopping foil placed at a distinct but moveable distance behind the target. In the first case, the energy of the emitted $\gamma$ ray is Doppler shifted. The decay probabilities of the populated states can be deduced from the intensity ratio of the shifted and the unshifted component of the corresponding transitions in the $\gamma$-ray energy spectrum. Using the Cologne Plunger device \cite{DEWALD2012786}, the distance between the target and the stopper foils can be varied in order to obtain the nuclear level lifetimes. In total 12 relative distances between 1\,µm and 1500\,µm were measured in approximately $165\,$hours of beam time. Distances were controlled by continuously measuring the voltage between the target and the stopper foils, hence, measuring the respective capacitance \cite{DEWALD2012786}.

\section{Data analysis}
\label{sec:data_analysis}
\begin{figure}[t]
    \centering
    \includegraphics[width=\linewidth]{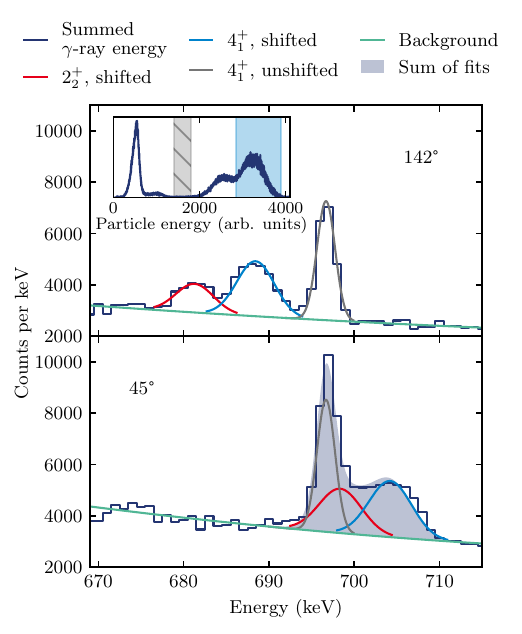}
    \caption{Comparison of $\gamma$-ray energy spectra at backward ($142$°) and forward ($45$°) angles summed over all distances. A coincidence condition was applied on the two-neutron transfer, i.e. the peak at high particle energies (blue area in the inset) with a background condition marked by the hatched area (shown in the inset). Both spectra are shown in an interval around the $4_1^+\rightarrow2_1^+$ transition at $696.7\,$keV with its Doppler-shifted components at $688.4$\,keV and $704.2$\,keV. In the backward spectrum the contaminant at $681.3\,$keV originating from the $2_2^+\rightarrow2_1^+$ transition is clearly visible. In the forward spectrum the contaminant is suspected to be located at $698.4\,$keV between the unshifted and the shifted component of the $4_1^+\rightarrow2_1^+$ transition. The deviation in events between forward and backward direction arises from the asymmetric detector setup with one HPGe more mounted at forward than at backward angles.}
    \label{fig:spectrum_697}
\end{figure}

The trigger-less data acquisition allows the definition of trigger conditions during the analysis of the data. The trigger was set to a one-$\gamma$-one-particle condition. To select the reaction channel of interest, i.e. the two-neutron transfer reaction, a coincidence condition on the corresponding high-energy structure of the particle spectrum was applied as shown in the inset of Fig. \ref{fig:spectrum_697}.

\begin{figure}[t]
    \centering
    \includegraphics[trim={1.5cm 1cm 1.5cm 3.5cm},width=\linewidth]{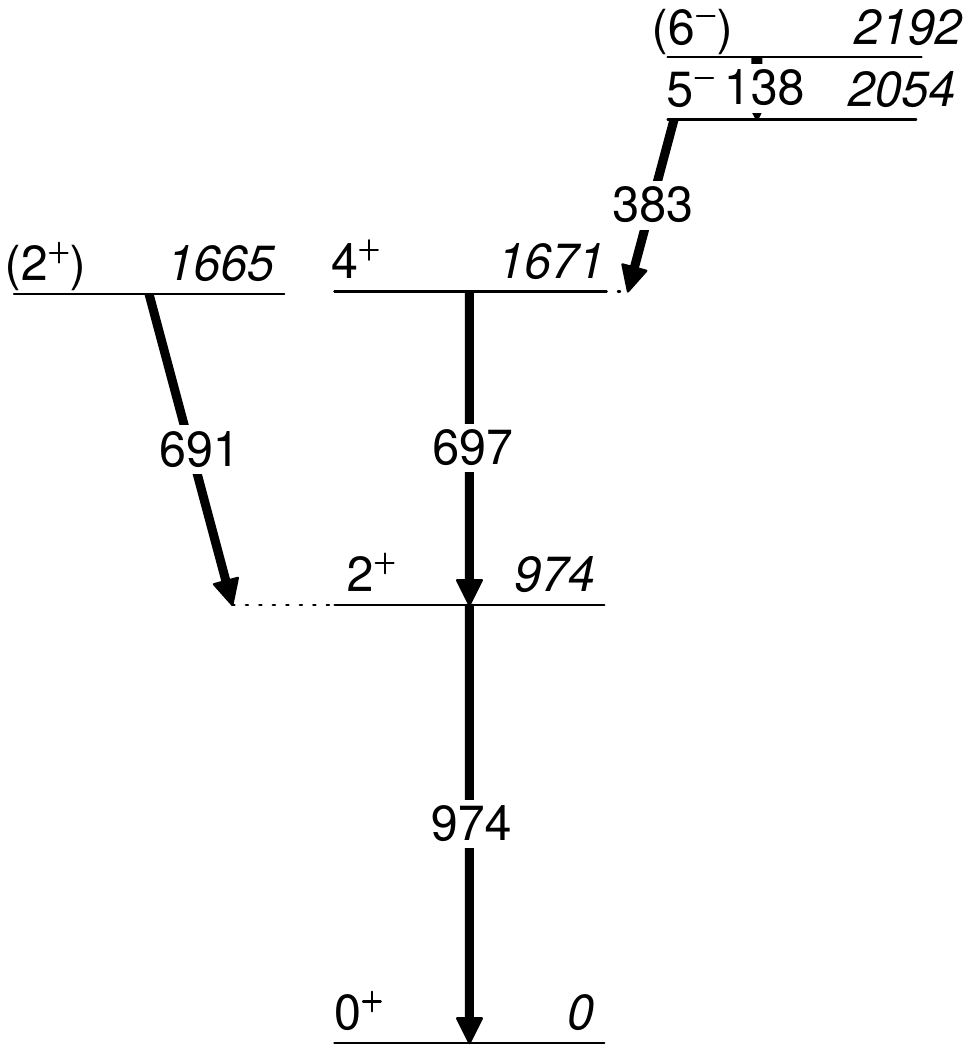}
    \caption{Level scheme of the five observed excited states of \Tett and their transitions that were populated in the experiment. Information on energy (keV), spin and parity quantum numbers is taken from Refs. \cite{A132} and \cite{Biswas}.}
    \label{fig:co_level_132te}
\end{figure}

The analysis was performed using the particle-gated $\gamma$ spectra. A particle-$\gamma\gamma$-coincidence analysis was not feasible. Therefore, feeding from higher-lying states into the states of interest had to be considered.
If excited states above the state of interest have a non-negligible lifetime relative to that of the state of interest, the ratio of the unshifted and shifted components of the state of interest is altered by the feeder. This leads to the measurement of a lifetime larger than the actual lifetime for the state of interest. Due to the specific mechanism of transfer reactions, they only populate states within a low energy and spin window, in contrast to fusion-evaporation reactions. Therefore, in the present analysis, it can be assumed that slow feeding contributions to the effective lifetime of the state of interest originate solely from the decays of few populated higher-lying states visible in the $\gamma$-ray spectrum, as suggested in \cite{PhysRevLett.104.042701}. In this case there are  no contributions from unobserved side feedings and the lifetime of interest can be corrected if the lifetimes of the feeding states are known. As depicted in the partial level scheme of $^{132}$Te in Fig. \ref{fig:co_level_132te}, observed feeding states are located at $2054\,$keV and $2192\,$kev, sequentially decaying into the $4^+_1$ state of $^{132}$Te via $\gamma$-ray transition energies of $383\,$keV and $138\,$keV, respectively. In the case of the tentative $6^-$ state at $2192\,$keV statistics are not sufficient to obtain a lifetime. However, it is assumed that the $6^-$ state is not directly feeding the $4^+_1$ state, because it would require an $M2$ transition instead of a more likely $M1$ transition to the $5^-$ state. The effective lifetime of the $5^-$ state contributes to the determination of the lifetime of the $4_1^+$ state. This effective lifetime of the $5^-$ state can be obtained from data, hence, accounting for both feeding contributions. 

The effective lifetime of the feeding $5^-$ state was obtained with the differential decay-curve method (DDCM) \cite{DDCM_Dewald1989} using the software \texttt{napatau} \cite{napatau}. A detailed description of the DDCM can be found in \cite{DEWALD2012786}. A spline of second order polynomials was fitted to the intensities of the shifted components. Simultaneously, the derivative of the polynomial was fitted to the intensities of the unshifted components. Including the uncertainty of the fit and the uncertainty of the particle velocity, the effective lifetime of the $5^-$ state was independently determined to be \mbox{$\tau_\text{eff, bwd}\left(5^-\right)=338(12)\,\text{ps}$} for the backward detectors, and \mbox{$\tau_\text{eff, fwd}\left(5^-\right)=341(13)\,\text{ps}$} for the forward detectors. Combining the values from both angles results in an effective mean lifetime of the $5^-$ state of
\begin{equation}
    \label{eq:lifetime_feeder}
    \tau_\text{eff}\left(5^-\right) = 339(9)\,\text{ps}.
\end{equation} 
The effective population of the $5^-$ state amounts to about 50\,\% of the population of the $4^+_1$ state as it is determined from the efficiency-corrected $\gamma$-ray intensities of their corresponding decay transitions.

Particle-gated $\gamma$-ray spectra in the region of interest for the $4_1^+\rightarrow2_1^+$ transition are displayed in Fig. \ref{fig:spectrum_697}, showing the spectra for both detector angles summed over all distances. The spectra show the unshifted and the shifted components of the $4_1^+\rightarrow2_1^+$ transition at $697\,$keV and a contaminant, which is best visible at backward angles at about $680\,$keV. This transition is identified as the Doppler-shifted component of the $2_2^+\rightarrow2_1^+$ transition at $691\,$keV. It is fully Doppler shifted due to the short lifetime of $\tau(2_2^+)=0.92(7)\,$ps \cite{tstetz} of the decaying one-phonon mixed-symmetry state at $1671\,$keV. Accordingly, no unshifted component of this transition is observed.

The unshifted and Doppler-shifted components of the $4^+_1 \rightarrow 2^+_1$ transition and the above-mentioned Doppler-shifted $2^+_2 \rightarrow 2^+_1$ transition, were fitted with Gaussian distributions using the analysis package \texttt{hdtv} \cite{hdtv}. The best parameters (mean energies and widths) for all peaks were determined from spectra, summed over all distances as shown in Fig. \ref{fig:spectrum_697}, and then kept constant to derive consistent fits for all distance settings. 

For the forward angles, where peaks maximally overlap, peak positions were fixed from calculating the respective Doppler shifts based on the observations at backward angles. The widths of the Doppler-shifted components are not equal in forward and backward angles to account for different detector responses of the two detector rings. Within the same angle, the widths of the Doppler-shifted $4_1^+\to2_1^+$ and $2_2^+\to2_1^+$ transitions are equal. For better control of possible systematics in the fits, in particular at forward angles, the lifetime analysis of the $4_1 ^+$ state was performed separately for the forward- and backward-angle detectors. In order to account for uncertainties of the fixed widths and positions, especially at forward angles, the fits were done several times with varying widths and positions within their uncertainty range.

An established approach to describe complex decay behaviour with feeding contributions are the so-called Bateman equations \cite{bateman}, using known lifetimes of the feeding states. A detailed description of the approach using the Bateman equations can be found in Ref.~\cite{DEWALD2012786}.
The intensities of the unshifted $I_\text{u}$ and the shifted $I_\text{s}$ components occurring in the $\gamma$-ray energy spectra are used to define the decay curve $R_i(t)$ for the excited state $i$:
\begin{equation}
    \label{eq:decay_curve}
    R_i(t) = \frac{I_i^\text{u}(t)}{I_i^\text{u}(t)+I_i^\text{s}(t)}\,.
\end{equation}
For an ensemble of $n$ nuclei in the state $i$ the Bateman equations are governed by the differential equation
\begin{equation}
    \frac{\mathrm{d}}{\mathrm{d}t}n_i(t) = -\lambda_in_i(t) + \sum_{k=i+1}^{N}\lambda_kn_k(t)b_{ki}\,,
\end{equation}
with the time $t$, feeding states $k$ with respective decay constants $\lambda_k$, and the branching ratios $b_{ki}$ of the feeding states $k$ to the state of interest $i$.
Using the exponential decay law and considering only one feeding state $k$ the solution of this differential equation is given by 
\begin{equation}
    \label{eq:bateman_final_decay}
    R_i(t)=P_ie^{-\lambda_it} + b_{ki}P_k \frac{\lambda_ie^{-\lambda_kt}-\lambda_ke^{-\lambda_it}}{\lambda_i-\lambda_k}\,,
\end{equation}
where $P_i, P_k$ are the population probabilities of the respective states. With a known feeder lifetime, the decay constant corresponding to the lifetime of the state of interest is the only free parameter and can be obtained by fitting Eq.~\eqref{eq:bateman_final_decay} to the experimental decay curve defined in Eq.~\eqref{eq:decay_curve} over all distances. The uncertainties of the lifetime fits were determined by probing their probability distributions by randomly varying the data points according to their individual uncertainties.

\begin{figure}[t]
    \centering
    \includegraphics[width=\linewidth]{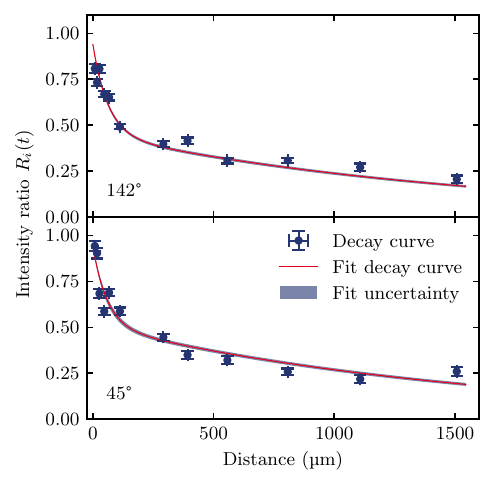}
    \caption{The fit of the Bateman equations to the intensity ratios at the backward (top) and forward (bottom) angle of the decay curve defined in Eq.~\eqref{eq:decay_curve} is shown. The lifetime of the $4^+_1$ state of \Tett is extracted from the fits because it is the only free fit parameter. The uncertainties introduced by varying peak widths and positions are not reflected in the depicted uncertainty bars but in the resulting values of the lifetime.}
    \label{fig:bateman}
\end{figure}

The fit to the observed intensity ratios [Eq.~\eqref{eq:decay_curve}] at backward angle, including the effective lifetime and population of the $5^-$ state, is shown in in the top panel of Fig. \ref{fig:bateman} and results in a lifetime of the $4^+_1$ state of $\tau_\text{bwd}(4_1^+) = 13.7(14)\,\text{ps}\,$. Similarly, the fit to the disentangled forward data is shown in the bottom of Fig. \ref{fig:bateman}, resulting in a lifetime of $\tau_\text{fwd}(4_1^+) = 12.8(21)\,\text{ps}\,$. The larger uncertainty reflects the larger peak-fitting uncertainties at forward angle. However, both values agree well within the uncertainty. The combined result is given by their weighted average of $\tau_{\text{av}}(4_1^+) = 13.4(12)\,\text{ps}$.

A possible contaminant from the $4_3^+ \rightarrow 4_1^+$ transition of \Teth\!, which is excited by the inelastic scattering of the $^{18}$O ions, cannot be excluded since the transition energy of $697.7\,$keV is very close to the transition of interest of \Tett at $697.1\,$keV. In addition, inelastic-scattering reactions (or Coulomb excitation close to the Coulomb barrier) cannot be resolved from two-neutron transfer reactions in the particle-energy spectrum in the present experiment. The contamination from the \Teth $4_3^+\to4_1^+$ transition can be estimated in the matrices without the applied particle-coincidence condition. In those matrices a $\gamma$-coincidence condition can be set on the $2_1^+\to0_1^+$ transitions once for \Teth and once for \Tett\!. Integration of the resulting structure around $697\,$keV, considering the amount of events that was gated on and rescaling to the particle-gated spectra, allows to quantify the level of contamination as 1.63(14)\%. Furthermore, the short lifetime of $\tau(4_3^+)=0.86^{+0.20}_{-0.14}\,$ps \cite{hicks2022} suggests that the \Teth contaminant only contributes to the shifted component of the transition of interest. We therefore assume the maximum impact of the contaminant by subtracting 1.63\% of the total intensity of the transition of interest from its shifted components for each distance. The near-negligible correction introduces an additional systematic uncertainty of $0.7\,$ps which is added in quadrature to $\tau_\text{av}$, leading to the final result for the lifetime of the $4^+_1$ state of \Tett from this experiment,
\begin{equation}
    \label{eq:final_lifetime}
    \tau(4_1^+) = 13.4(14)\,\text{ps}\,.
\end{equation}

\section{Discussion}
With the usual assumption of pure $E2$ character for the $4^+_1 \rightarrow 2^+_1$ transition, and a conversion coefficient of $\alpha=3.372\cdot10^{-3}$ \cite{bricc}, the lifetime from Eq.~\eqref{eq:final_lifetime} yields the $E2$ transition strength:
\begin{equation}
    B(E2; 4^+_1\rightarrow 2^+_1) = 370(39)\, \text{e}^2\text{fm}^4 = 9.3(10)\, \text{W.u.}
\end{equation}

\begin{table}[b]
    \caption{Lifetime of the $4_1^+$ state of $^{132}$Te and the deduced $B(E2; 4^+_1\rightarrow 2^+_1)$ value obtained in this work are listed and compared to corresponding data from a previous fast-timing measurement \cite{Kumar} and current shell-model calculations.}
    \label{tab:lifetimes}
    \begin{tabular}{llll}
    \hline
    \hline
    \noalign{\vskip 2pt}
        & $\tau$ (ps) & $B(E2)$ (e$^2$fm$^4$) & $B(E2)$ (W.u.)\\
        \hline
        \noalign{\vskip 2pt}
        Kumar \textit{et al.} &  61(5) & \phantom{0}80(8) & 2.0(2)\\
        This work & 13.4(14) & 370(39) &9.3(10) \\
        Shell model & & 324 & 8.1\\
    \hline
    \hline
    \end{tabular}
\end{table}

\begin{figure}[t]
    \includegraphics[width=\linewidth]{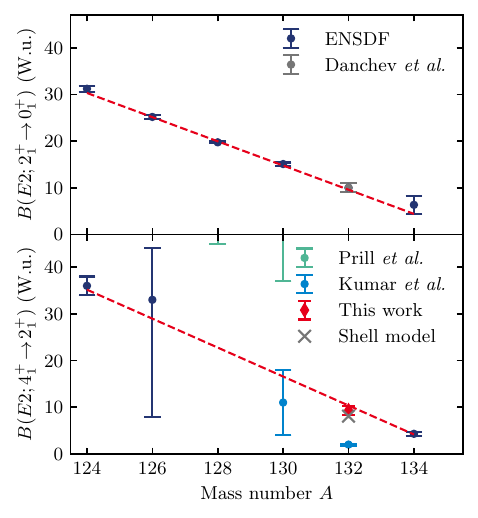}
    \caption{$E2$ transition rates for different tellurium isotopes as shown in Fig.~\ref{fig:b(e2)_evol}. A linear function was fitted to the experimental $E2$ strengths of the $2^+_1\to0_1^+$ (top) and the $4^+_1\to2_1^+$ (bottom) transitions. The linear functions show a similar slope with $\Delta B(E2; 2_1^+\to0_1^+)/\Delta N\!= -2.59(9)\,$W.u. and $\Delta B(E2; 4_1^+\to2_1^+)/\Delta N\!= -3.10(19)\,$W.u. The data from Prill \textit{et al.}~\cite{prill} and Kumar \textit{et al.}~\cite{Kumar} were not included in the fit.}
    \label{fig:4+_evol}
\end{figure}

This value is shown in Fig.~\ref{fig:4+_evol}, along with literature values for $A=124-134$ Te isotopes. We consider the values for $^{128,130}$Te from Prill \textit{et al.}~\cite{prill} as inaccurate and misleading. In addition, their data point on $^{128}$Te has an uncertainty such that it cannot contribute to the following discussion about the structural configuration of the respective $4^+_1$ states in a relevant way. While they are indicated in Fig.~\ref{fig:4+_evol}, they are not further considered in the following discussion. For $^{130}$Te, values from Kumar \textit{et al.\!}~\cite{Kumar} and Prill \textit{et al.\!}~\cite{prill} disagree beyond their reported uncertainties; the former in support of little, if any, collectivity and the latter suggesting enhanced collectivity in the $4^+_1$ state. 

For $^{132}$Te, the present isotope of interest, our measurement of the $B(E2;4^+_1 \rightarrow 2^+_1)$ value rules out the value claimed by Kumar \textit{et al.\!} \cite{Kumar}. The RDDS technique is well-applicable in the relevant lifetime range around 10\,ps, while the fast-timing technique used for the measurement by Kumar \textit{et al.\!} in Ref.~\cite{Kumar} reaches the limit of its applicability or is even beyond. Therefore, while their reported value is still displayed in Fig.~\ref{fig:4+_evol}, it is not further considered for the quantitative analysis in our subsequent discussion. 

Considering the remaining literature $B(E2)$ values, a linearly declining trend of $B(E2; 4^+_1\rightarrow 2^+_1)$ values as a function of neutron number towards $N=82$ appears to develop. It has previously been discussed in the framework of the Interacting Boson Model-2 by Sambataro~\cite{SAMBATARO1982365}. The trend of the $B(E2; 4^+_1\rightarrow 2^+_1)$ values is akin to the linear trend of the $B(E2;2^+_1\rightarrow 0^+_1)$ values which are included in the top panel of Fig.~\ref{fig:4+_evol}. Such a trend would be typical for collective vibrators. However, in the collective vibrational limit a factor of two difference in the slopes would be expected if using the same effective charges for both transition strengths. In contrast, in tellurium isotopes both slopes are rather similar. 
Their ratio, 
\begin{equation}
\frac{\Delta B(E2; 4^+_1\to2_1^+)/\Delta N}{\Delta B(E2; 2^+_1\to0_1^+) / \Delta N} = 1.20(9)   
\end{equation}
may be considered as a \textit{mean $B_{4/2}$ ratio}, averaged over the corresponding part of the isotopic chain. It is less sensitive to local peculiarities of the underlying single-particle structure than a local value for a selected nuclide. Its value above unity suggests that the lowest excited states of the heavy tellurium isotopes may not simply be dominated by seniority. Quadrupole collective components may contribute to their wave functions that, however, do not yet dominate due to the proximity of the tellurium isotopic chain to the $Z=50$ shell closure. This aspect will be further discussed below in terms of the shell model. 

%The $B_{4/2}$ ratio of $^{130}$Te using the data from \cite{prill,A130} is $B_{4/2}(^{130}{\rm Te}) = 3.1(7)$. Considering its large uncertainty, it is in fair agreement with a collective (vibrational) structure. Therefore, $^{132}$Te can be expected to be at the intersection between the more collective structure of $^{130}$Te and the closed-shell seniority-type structure of $^{134}$Te.

%\begin{figure}[t]
%    \includegraphics[width=\linewidth]{pictures/BE2_evolution_Te_inset.pdf}
%    \caption{Experimental $B(E2; 4_1^+\rightarrow2_1^+)$ values for different tellurium isotopes as shown in Fig.~\ref{fig:b(e2)_evol}. The inset enlarges the values for $^{132}$Te and $^{134}$Te, including the result from this work an the shell-model calculation. The values of $^{124}$Te, $^{126}$Te and $^{134}$Te are taken from Refs.~\cite{Saxena024316,STOKSTAD1967507,A134}, respectively.}
%    \label{fig:4+_evol}
%\end{figure}

\begin{figure*}[t]
    \includegraphics[width=\linewidth]{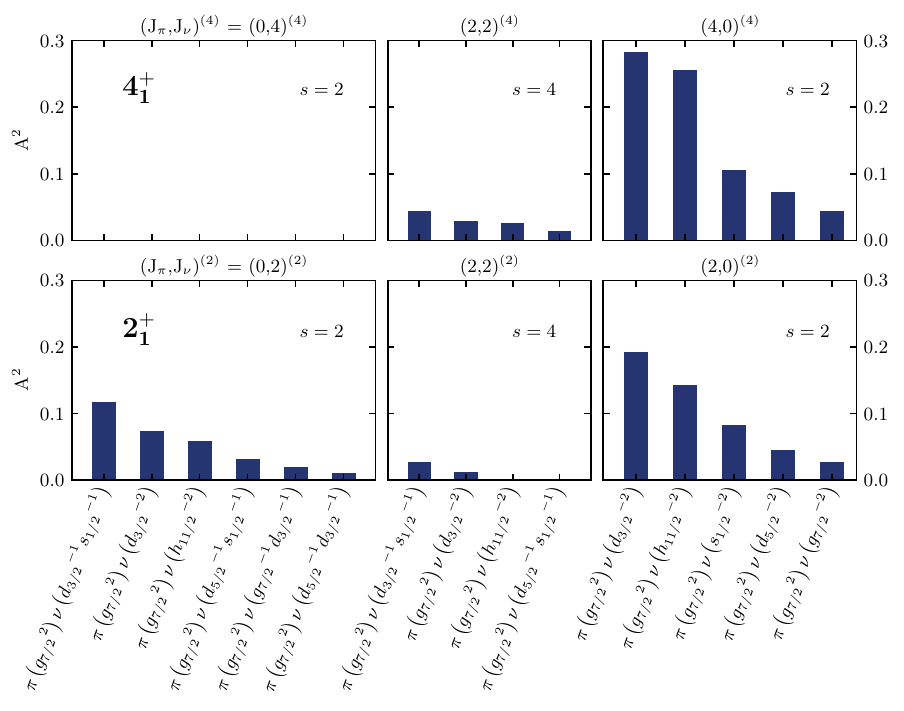}
    \caption{The calculated squared amplitudes of configurations which contribute more than 1\% to the total wave function of the $4_1^+$ (top) and the $2_1^+$ (bottom) states are shown. On the left are the seniority-like $s=2$ configurations where the valence protons couple to zero, whereas on the right the seniority-like $s=2$ configurations where the neutrons couple to zero are found. The middle panels amplitudes of seniority $s=4$ configurations are shown, where both, valence protons and neutrons, couple to the total angular momentum. In the case of the $4_1^+$ state they contribute with 11\% to the total wave function. In sum the shown squared amplitudes with $(J_\pi,J_\nu)^{(4)}=(4,0)^{(4)}$ contribute 76\% to the total wave function of the $4_1^+$ state.}
    \label{fig:amplitudes}
\end{figure*}

As mentioned in the introduction, the $R_{4/2} = 1.72$ ratio of $^{132}$Te is significantly below the vibrational limit of $R_{4/2}^\text{vib} = 2$. This is rather typical for nuclei in the close vicinity of a shell closure. Seniority-type configurations may be strongly present in either states' wave function, however, excited states even on closed shells with a sufficient amount of valence nucleons can have competing collective configurations. A prominent example for this behaviour is found in the tin isotopes at $Z=50$ \cite{PhysRevC.59.1930, BANDYOPADHYAY2005206}. The competition between $j^2$ and more complicated, collective configurations in $^{132}$Te is particularly interesting since this isotope has only two valence protons and neutrons, each. Hence, collectivity in particular in its $4^+_1$ state is at best just forming.

Besides the $R_{4/2}$ ratio, $E2$ strengths and the $B_{4/2}$ ratio are direct indicators of collectivity in the wave functions of the parent and daughter states of the decays. %The lifetime of the $4^+_1$ state measured in the present work is about a factor of five smaller than the previous literature value of Ref.~\cite{Kumar}. Accordingly, we obtain a larger $B(E2; 4^+_1 \rightarrow 2^+_1)$ value. 
Using the literature value of $B(E2;2^+_1 \rightarrow 0^+_1) = 10(1)\,$W.u.~\cite{Danchev}, we obtain a local $B_{4/2} = 0.93(13)$ value for $^{132}$Te. 
It is well below the collective limits, especially below the spherical-vibrator limit of $B_{4/2}=2$. Nevertheless, the absolute $B(E2)$ values of transitions depopulating the $2^+_1$ and $4^+_1$ states are both around 10 W.u., which is pointing to early signs of emerging collective behaviour.

In order to gain insight into the microscopic structure of the $4^+_1$ state and the slight enhancement of its $E2$ strength with respect to the expectation for a small, seniority-conserving $B(E2; 4^+_1 \rightarrow 2^+_1)$ value of about 1-2 W.u. for a seniority-dominated state, we analyse results from shell-model calculations. The same calculations had been employed for the previous study of low-lying states of $^{132}$Te from a Doppler-shift attenuation measurement with focus on the mixed-symmetric one-phonon $2^+_\text{ms}$ state of Ref.~\cite{tstetz}. The SN100PN interaction \cite{BrownSN100PN} and the code KSHELL \cite{kshell1,kshell2} have been employed to perform the calculations in a valence space composed of the $0g_{7/2}$, $1d_{5/2}$, $1d_{3/2}$, $2s_{1/2}$, and $0h_{11/2}$ orbitals for both, protons and neutrons. Proton and neutron effective charges were set to $e_\pi$\,=\,1.7\,$e$ and $e_\nu$\,=\,0.8\,$e$, which have widely been used in this region (see, e.g. Refs. \cite{gray2020,hicks2022}). The calculations predict a $B(E2; 4^+_1 \rightarrow 2^+_1)$ value of 8.1\,W.u., in rather good agreement with the present result, just below the one-sigma uncertainty limit. The discussed experimental and theoretical $B(E2; 4^+_1 \rightarrow 2^+_1)$ values for the $4^+_1 \rightarrow 2^+_1$ transition of $^{132}$Te are summarised in Table \ref{tab:lifetimes}. 

Akin to Ref. \cite{tstetz} we performed an analysis of the $4^+_1$ state's wave function, decomposing it into its basis configurations. All leading configurations with squared amplitudes $A^2 > 1$\% are considered. The amplitudes are illustrated in Fig.~\ref{fig:amplitudes}, which is to be compared to Fig. 3 of Ref. \cite{tstetz} giving the leading amplitudes for $2^+$ states of $^{132}$Te. The wave function of the $4^+_1$ state is dominated by proton configurations, i.e. two protons in a given orbital coupling to angular momentum $J_{\pi} = 4$, $[\pi(j^2)]^{(4)}$, and neutrons coupling to $J_{\nu}=0$. Such $(J_\pi,J_{\nu})^{(4)}=(4,0)^{(4)}$ configurations are typical seniority-scheme excitations with seniority $s=2$. Interestingly, neutron $(J_\pi,J_{\nu})^{(4)}=(0,4)^{(4)}$ configurations do not significantly contribute to the wave function and are all below the 1\% limit. The underlying reason is likely the presence of the neutron $\nu(s_{1/2})$ and $\nu(d_{3/2})$ orbitals which are filled towards the end of the shell, and which cannot contribute to $J_{\nu}=4$ configurations. In contrast, such configurations, namely, $(J_\pi,J_{\nu})^{(2)}=(0,2)^{(2)}$, do play a role in the formation of the $2^+_1$ state, as shown in the bottom panels of Fig.~\ref{fig:amplitudes}.

In a valence space composed of only two valence neutrons and protons, a collective-vibrational configuration would be formed by $[\pi(j^2)^{(2)} \nu(j^2)^{(2)}]^{(4)} = (2,2)^{(4)}$ configurations, where the protons and neutrons pairwise couple to $J_{\pi,\nu}=2$, and then form the $J=4$ state. This corresponds to two quadrupole pairs coupling to a total angular momentum $J=4$. In total, such configurations contribute about 11\% to the wave function of the $4^+_1$ state (see top-center panel of Fig. \ref{fig:amplitudes}). The seniority-like $(J_\pi,J_\nu)^{(4)}=(4,0)^{(4)}$ configurations dominate the wave function with 76\%, but connect to the $2^+_1$ state mostly by seniority-conserving $E2$ matrix elements, as can be inferred from comparing the proton parts of the wave functions in Fig.~\ref{fig:amplitudes}. Even though there are several $E2$ matrix elements which contribute to the total transition strength, in essence it is a seniority-preserving transition in the $\pi(g_{7/2})$ orbital. No contribution from the neutron part of the wave functions can be expected due to the almost vanishing $(0,4)^{(4)}$ part in the structure of the $4^+_1$ state. Then, it is the smaller $(2,2)^{(4)}$ part (cf. the middle panels in Fig. \ref{fig:amplitudes}) of the wave function that is responsible for the slight increase in the degree of collectivity, as reflected in the measured $B(E2; 4_1^+\to2_1^+)$ value.

\section{Summary}
The lifetime of the $4^+_1$ state of $^{132}$Te has been measured in an RDDS experiment with a precision of about 10 \% resulting in $\tau(4_1^+) = 13.4(14)\,\text{ps}$. The corresponding $B(E2;4^+_1 \rightarrow 2^+_1)$ value of $9.3(10)\,$W.u. exceeds the previous literature value by about a factor of five and invalidates it. The present value agrees with predictions within the shell model and indicates signs for emerging collectivity in the structure of the $4^+_1$ state. The analysis of its shell-model wave function shows that its proton contribution is dominated by seniority $s=2$, spin $J=4$ configurations, whereas neutron contributions of this type to the wave function are small. 
Proton-neutron coupled contributions with seniority $s=4$ cause the enhancement of the $B(E2;4^+_1 \rightarrow 2^+_1)$ value over the seniority-conserving expectation and are consequently, responsible for the onset of collectivity in $^{132}$Te.

Processed data shown in the figures of this article are openly available at the TUdatalib repository of Technische Universität Darmstadt~\cite{tudatalib_132Te}. The raw data corresponding to the findings in this manuscript are not publicly available upon publication because it is not technically feasible and/or the cost of preparing, depositing, and hosting the data would be prohibitive within the terms of this research project. The raw data are available from the authors upon reasonable request.

\begin{acknowledgments}
    We thank the staff of the Cologne tandem accelerator for providing stable beam conditions throughout the experiment. This work was supported by the German BMFTR under Grant No 05P24RD3, BMBF under Grant Nos  05P21RDCI2, 05P21RDFN9, 05P19RDFN1 and 05P15PKFNA and by the German Research Foundation DFG as part of the Project-ID 264883531 – Research Training Group 2128 ’Accelence’, Project-ID 499256822 – Research Training Group 2891 ’Nuclear Photonics’, JO 391/20-1, FR 3276/2-1 and DE 1516/5-1. We further acknowledge the DFG for the upgrade of the used Germanium detectors under grant INST 216/988-1 FUGG. K.G, D.K, and G.R acknowledge the support by the European Union-NextGenerationEU, through the National Recovery and Resilience Plan of the Republic of Bulgaria, project No BG-RRP-2.004-0008-C01, by DAAD under the partnership agreement between the University of Cologne and University of Sofia and by the Bulgarian Ministry of Education and Science, within the National Roadmap for Research Infrastructures (object CERN).
\end{acknowledgments}

\bibliography{library}

\end{document}